\documentclass{article}

\pdfoutput=1

\usepackage{cite}
\usepackage{graphicx}
\usepackage{caption}
\usepackage{amssymb}
\usepackage{array}
\usepackage{dcolumn}
\usepackage{bm}
\usepackage{textgreek}
\usepackage{amsmath}
\usepackage[letterpaper, total={7in, 9in}]{geometry}
\usepackage{authblk}

\title{Verniered Optical Phased Arrays for Grating Lobe Suppression and Extended FOV}

\author[1,*]{Nathan Dostart}
\author[2]{Bohan Zhang}
\author[1]{Michael Brand}
\author[1]{Daniel Feldkhun}
\author[2]{Milo\v s Popovi\'c}
\author[1]{Kelvin Wagner}

\affil[1]{Department of Electrical, Computer, and Energy Engineering, University of Colorado, Boulder, CO, 80309, USA}
\affil[2]{Department of Electrical and Computer Engineering, Boston University, Boston, MA, 02215, USA}

\affil[*]{Corresponding author: nathan.dostart@colorado.edu}

\date{\today}

\begin{document}
	
\maketitle

\begin{abstract} 
Optical phased arrays (OPAs) which beam-steer in 2D have so far been unable to pack emitting elements at $\lambda/2$ spacing, leading to grating lobes which limit the field-of-view, introduce signal ambiguity, and reduce optical efficiency. Vernier schemes, which use paired transmitter and receiver phased arrays with different periodicity, deliberately misalign the transmission and receive patterns so that only a single pairing of transmit/receive lobes permit a signal to be detected. A pair of OPAs designed to exploit this effect thereby effectively suppress the effects of grating lobes and recover the system's field-of-view, avoid potential ambiguities, and reduce excess noise. Here we analytically evaluate Vernier schemes with arbitrary phase control to find optimal configurations, as well as elucidate the manner in which a Vernier scheme can recover the full field-of-view. We present the first experimental implementation of a Vernier scheme and demonstrate grating lobe suppression using a pair of 2D wavelength-steered OPAs. These results present a route forward for addressing the pervasive issue of grating lobes, significantly alleviating the need for dense emitter pitches.
\end{abstract}

\section{Introduction}
\label{sec:introduction}

Optical phased arrays (OPAs), which use an array of emitting elements to project (or receive from) a controlled illumination pattern, are of current interest to the academic and industrial communities due to the ever-increasing demand for smaller, lighter, and more energy-efficient devices for communications and sensing. Integrated photonic OPAs have been a particular focus of recent research efforts due to the promise of dense OPA designs, agile beam steering, and co-integration with advanced electronics. Additionally, with full phase and amplitude control \cite{abediasl2015monolithic}, or 2D wavelength-steered designs \cite{van2011two,dostart2020serpentine}, OPAs can emit multiple independently controlled beams simultaneously. These OPA beam-steering systems have been used for applications such as free-space communication links \cite{poulton2019long}, imaging systems \cite{aflatouni2015nanophotonic,raval2016nanophotonic,fatemi2018high,clevenson2020incoherent}, or LIDAR \cite{poulton2019long,dostart2019vernier}.

\begin{figure*}[t]
	\centering
	\includegraphics[width=\textwidth]{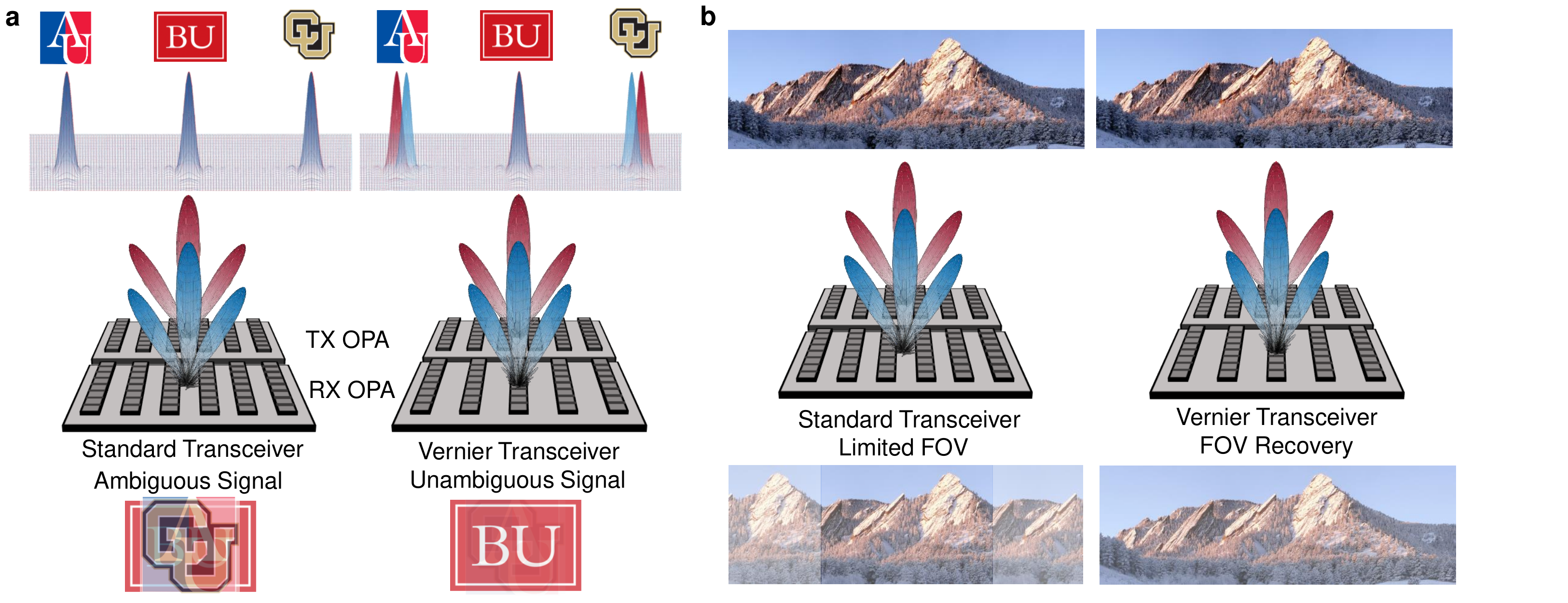}
	\caption{\textbf{Vernier transceiver as compared to a `standard' OPA transceiver using identical transmit (TX) and receive (RX) OPAs.}
	\textbf{a} Reduction of signal ambiguity: misaligned grating lobes ensure a transmitted signal can only be received from a single lobe, resulting in disambiguation of the detected signal.
	\textbf{b} Increased FOV: whereas a `standard' OPA transceiver has a FOV limited by the grating lobe spacing, the single lobe alignment allows a Vernier transceiver to achieve of FOV limited only the radiation pattern of a single OPA element. `Flatirons Winter Sunrise' by Jesse Varner and AzaToth is licensed under CC BY-SA 2.5, https://en.wikipedia.org/wiki/Flatirons.}
	\label{fig:concept}
\end{figure*}

While integrated photonic OPAs enable significant improvements over bulk-optic beam-steering, they also suffer from grating lobes. Grating lobes are undesired beams which mirror the main (desired) beam, arising in OPAs with element pitches larger than $\lambda_0/2$. Even high-index platforms such as silicon have not been able achieve the waveguide pitch required for grating lobe-free operation of 2D beam steering without significant cross-talk between adjacent waveguides. To our knowledge, the densest pitch achieved so far for 2D beam steering is 1.3 {\textmu}m \cite{zhang2019sub,miller2020large}. For most applications such as LIDAR, these grating lobes limit the field-of-view (FOV) of the system to the grating lobe spacing, introduce spurious signals, and reduce optical power emitted into the main lobe. The spurious signals, resulting from back-scattered light excited from/received by the grating lobes, is a particularly grievous issue that cannot be compensated for post-measurement \cite{dostart2020serpentine}. Additionally, limitations on FOV generally restrict to less than the oft-quoted 70$^\circ$ desired for automotive LIDAR applications; in most cases a much smaller FOV is achieved. An approach to alleviating both of these issues, spurious signals and limited FOV, would therefore benefit all OPAs immensely.

Several approaches to suppressing grating lobes have been proposed or demonstrated in OPAs, specifically sparse arrays \cite{hutchison2016high,komljenovic2017sparse,fatemi2019nonuniform}, Vernier arrays \cite{pinna2018vernier}, and element pitches below $\lambda_0/2$ \cite{phare2018silicon}. Simply avoiding grating lobes entirely, and avoiding the associated power loss, is clearly the ideal solution. A recent paper \cite{phare2018silicon} has demonstrated this high-density pitch by varying waveguide widths in order to suppress adjacent waveguide cross-talk and achieved grating lobe-free operation. However, using this approach for 2D beam steering likely requires significant limitations on both grating length and operation bandwidth to avoid significant cross-talk, leaving the issue of grating lobes an open question. The preferred approach of several groups to avoiding grating lobes has been to use aperiodic or `random' arrays to avoid the periodicity that gives rise to grating lobes \cite{hutchison2016high,komljenovic2017sparse,fatemi2019nonuniform}. Such an approach spreads the power in the grating lobes across the entire FOV of the OPA, forming only one beam and recovering the full FOV. However, this power radiated to all angles can still create spurious signals when back-scattered that will reduce the fidelity of an imaging or LIDAR system. The Vernier approach, which has been recently proposed \cite{pinna2018vernier}, can both recover the full FOV and avoid these spurious signals by co-design of a pair of transmitter (TX) and receiver (RX) OPAs. The Vernier transceiver, which can only transmit and receive along a single, aligned pair of lobes, thereby effectively suppresses grating lobes at the system level.

In this work, we discuss the Vernier approach and find a Vernier transceiver configuration which optimally suppresses spurious signals while recovering the full FOV. We consider a common 2D OPA geometry and analytically examine the relation between array pitch and grating lobe suppression, and determine the phase functions required to achieve full FOV recovery without ambiguity. We then consider the geometry of our example implementation, a serpentine optical phased array (SOPA) which uses wavelength-steering along both dimensions \cite{dostart2020serpentine}, and derive the design constraints for the Vernier conditions. We experimentally demonstrate grating lobe suppression for improved SNR and reduced ambiguity in a pair of SOPAs, the first experimental demonstration of the Vernier transceiver approach.

\section{Vernier Transceiver Concept}
\label{sec:concept}

The improved performance of a Vernier OPA transceiver over the `standard' transceiver design, using identical TX and RX OPAs, is shown in Fig.~\ref{fig:concept}. The standard transceiver (center, left) uses periodic OPAs, here with both TX and RX OPAs having 6 gratings. The identical periodicity of both OPAs results in a far-field set of grating lobes (top) which is identical for both the transmitted intensity (red) and reception pattern (blue). A Vernier transceiver (center, right) has different grating pitches, here using the same TX OPA with 6 gratings whereas the RX OPA has 5 gratings. This results in two different far-field patterns, where for proper grating phases the TX and RX main (center) lobes are aligned where as their grating lobes to either side are not aligned.

One capability of a Vernier transceiver is to reduce the ambiguity created by grating lobes, shown schematically in Fig.~\ref{fig:concept}\textbf{a}. A transceiver using identical TX and RX OPAs (left) projects a set of overlapping lobes into the far-field. A transmitted pulse would hit every location within the red lobes, scatter off the target, and can be received by the blue lobes. For our example targets, AU/BU/CU, this creates an ambiguous signal. This ambiguity is shown schematically at the bottom for the case of imaging these targets, where even though more power is received along the main lobe it is difficult to distinguish the desired target. This contrasts with the Vernier transceiver case, where the main lobes of the TX and RX OPAs overlap but the grating lobes are misaligned. This results in significantly less signal received from the targets sampled by the grating lobes, and a far less ambiguous signal.

The other desired capability of a Vernier transceiver is to increase the FOV beyond the grating lobe spacing, which limits the FOV for a standard transceiver. This capability is shown in Fig.~\ref{fig:concept}\textbf{b} (left), where only the central portion of the scene can be imaged with a standard transceiver (the grating lobe-limited FOV). When attempting to image the scene past the limited FOV, the a lobe (which was previously a grating lobe) scrolls into the central FOV and becomes the effective main lobe, contributing a stronger signal than the lobe directed outside the central FOV. A Vernier transceiver (right) allows the same lobe to be used across the entire FOV due to the lobe alignment, ensuring the desired lobe is always distinguished from the other lobes.

\section{Vernier Scheme Theoretical Analysis}
\label{sec:theory}

\begin{figure*}[t]
	\centering
	\includegraphics[width=\textwidth]{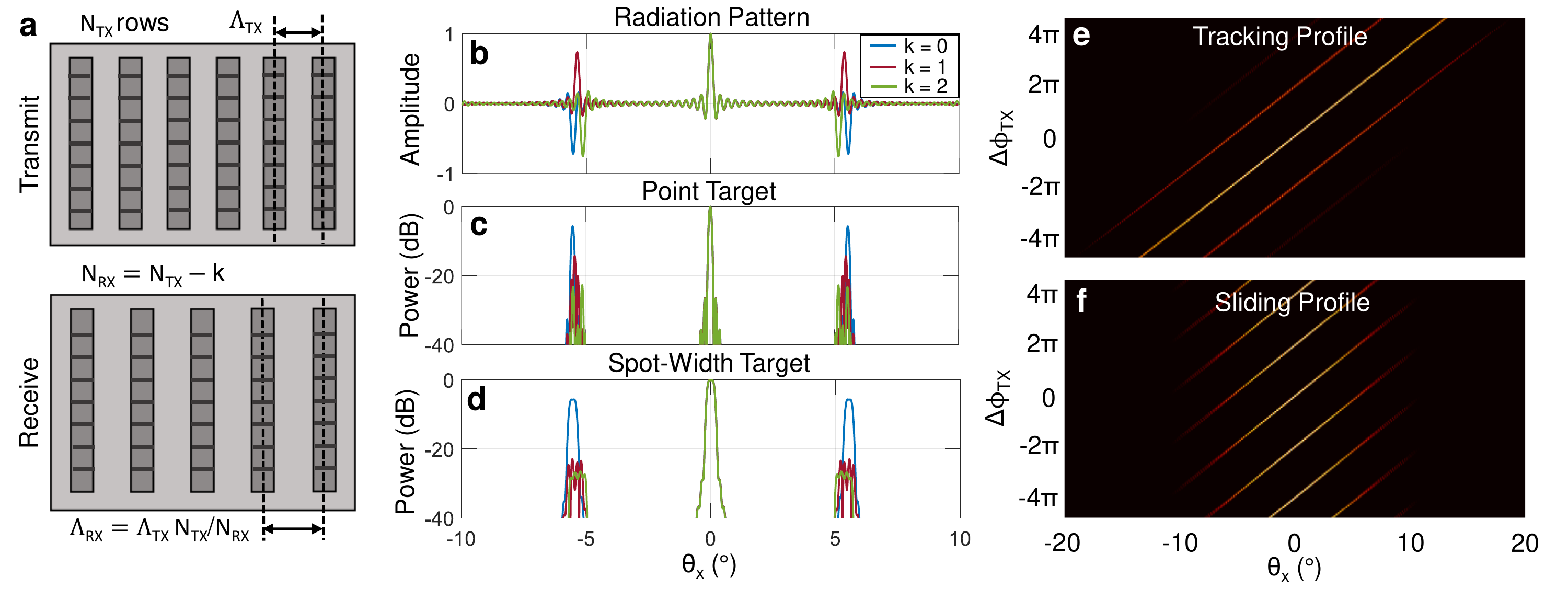}
	\caption{\textbf{Transceiver geometry for grating lobe suppression and FOV recovery.}
	\textbf{a} Proposed transceiver geometry, where the widths of both TX (top) and RX (bottom) OPAs are identical but the RX OPA has $k$ fewer gratings.
	\textbf{b} Far-field radiation pattern for three RX OPA designs with $k=0$ (blue), $k=1$ (red), and $k=2$ (green).
	\textbf{c} Detector signal for a point target; the Vernier transceivers ($k=1$, $k=2$) have 8.6 dB and 17.7 dB grating lobe suppression relative to a standard ($k=0$) transceiver.
	\textbf{d} Detector signal for a target which is the same width as the grating lobe; the Vernier transceivers have 17 dB and 20.8 dB grating lobe suppression.
	\textbf{e} `Tracking' steering mode for $k=1$: the overlapped TX and RX radiation patterns (the signal shown in \textbf{c}, red) plotted vs. the grating-to-grating phase of the TX OPA ($\Delta\phi_{TX}$). The overlapping main lobes are scanned across more $20^\circ$, significantly larger than the $5.5^\circ$ grating lobe spacing.
	\textbf{f} `Sliding' steering mode for $k=1$: the overlapping lobes are constrained to lie within the central FOV.
	}
	\label{fig:theory}
\end{figure*}

We begin by analytically examining the most common geometry for integrated 2D OPAs: a 1D array of long, weak gratings (see for example \cite{hutchison2016high,poulton2019long}). Wavelength-steering is used to control the emission angle along the gratings by using the dispersive nature of grating couplers. Along the orthogonal dimension, beam steering is achieved by preceding each grating with a phase shifter which allows arbitrary control of the gratings' emission phases. We consider the case of designing an RX OPA for some given TX OPA, where we desire to maximally suppress the grating lobes using the Vernier effect. In this regard we will present the derivation from an `intuitive' perspective; a full derivation is provided in the Supplemental Materials which arrives at the same conclusions.

Examination of the far-field patterns of both the TX and RX, and use of the reciprocity principle, allows us to calculate the optical efficiency for different angles. We calculate the improvement in ambiguity, SNR, and scan range of the Vernier arrangement by comparison to two reference cases: a standard incoherent detector, and identical TX/RX arrays. Additionally, we consider two distinct phase-steering functions, a `naive' phase function and a FOV-recovering phase function, and demonstrate the latter allows for a full semi-circle FOV which, until now, has been limited to the grating lobe pitch.

As a final theory component, we address the question of recreating the optimal RX array conditions and phase steering functions in a 2D wavelength-steered OPA. Because the geometry along the Vernier dimension is \emph{not} decoupled from the phase of each grating, the implementation of the phase steering functions in a 2D wavelength-steered OPA is non-trivial.

\subsection{Grating Lobe Suppression with a Vernier Transceiver}
\label{subsec:rx_design}

Here we consider two 2D OPAs used as the TX and RX ends of a transceiver with $N_{TX/RX}$ gratings of width $w$ and spaced with pitch $\Lambda_{TX/RX}$ forming apertures of widths $W_{TX/RX} = N_{TX/RX}\Lambda_{TX/RX}$. For single-mode operation of the gratings, and negligible cross-talk between adjacent gratings, the aperture emission is separable and therefore we consider the 1-D case along the grating-orthogonal direction, denoted here as the $x$-axis. Furthermore, we assume monochromatic input $\lambda$ and excitation only of the in-plane electric field component; therefore we leave the polarization implicit and use the scalar wave equation approximation \cite{goodman2005introduction}.

For simplicity we assume uniform emission across the width of the grating, such that the field at the grating interface can be accurately described by a normal plane wave impinging on a rectangular aperture with width $w$. While the mode profile or full-field simulations can be used to calculate more accurate emission profiles, they do not affect grating lobe suppression (relative to the identical TX/RX case). Each grating is assumed to have an arbitrary (controlled) phase and identical amplitude, corresponding to a uniformly apodized aperture along the $x$-dimension. It should be noted that most common apodization strategies, such as windowed Gaussians, will only \emph{increase} sidelobe and grating lobe suppression, so we consider the `worst-case' scenario of uniform/no apodization.

In most applications, both the number of gratings, array size, and grating pitch of the TX array have been limited by some external factors, e.g. number of controllable phase-shifters and minimum waveguide pitch to avoid cross-talk. We therefore restrict the RX array such that $\Lambda_{RX}\geq\Lambda_{TX}$, $N_{RX}\leq N_{TX}$, and $W_{RX}\leq W_{TX}$. 

We begin with the scalar effective aperture function for the TX array

\begin{equation}
    U_{TX}(x) = \text{rect}\left(\frac{x}{W_{TX}}\right)\left[\text{rect}\left(\frac{x}{w}\right)\ast\text{comb}\left(\frac{x}{\Lambda_{TX}}\right)\right].
\end{equation}

Because we are interested in the angular distribution of the light to find the far-field radiation pattern, we take the spatial Fourier transform of the aperture which yields

\begin{equation}
    F_{TX}(f_x) = \text{sinc}\left(W_{TX}f_x\right)\ast\left(\text{sinc}(wf_x)\text{comb}\left(\Lambda_{TX}f_x\right)\right)
\end{equation}

\noindent where we have used the definition $\text{sinc}(x)=\sin{\pi x}/(\pi x)$, dropped normalization factors, and identified $f_x$ as the spatial frequency of the Fourier transform.

We now use the principle of reciprocity \cite{haus1984waves} to identify the field which will be $100\%$ coupled into the RX array, a coherent receiver, as the field which the RX array would radiate if light is injected into the output of the array \cite{dostart2020serpentine}. The RX array has similar effective aperture and angular distribution of radiation as the TX array with appropriate exchanging of subscripted variables

\begin{equation}
    U_{RX}(x) = \text{rect}\left(\frac{x}{W_{RX}}\right)\left[\text{rect}\left(\frac{x}{w}\right)\ast\text{comb}\left(\frac{x}{\Lambda_{RX}}\right)\right]
\end{equation}
\begin{equation}
    F_{RX}(f_x) = \text{sinc}\left(W_{RX}f_x\right)\ast\left(\text{sinc}(wf_x)\text{comb}\left(\Lambda_{RX}f_x\right)\right).
\end{equation}

In the far-field regime, at angles near broadside (the direction normal to the chip plane), the Fraunhofer approximation can be used for both TX and RX arrays to convert the spatial frequency to a spatial coordinate on a plane at distance $z$ as  $f_x=x'/\lambda z$. For a diffuse target in this plane with field reflectivity profile $\mathcal{R}(x')$, the power received by the RX array can be written as

\begin{equation}
    P_{det} \propto \left|\int_{x'}dx'\mathcal{R}(x')F_{TX}\left(\frac{x'}{\lambda z}\right)F_{RX}\left(\frac{x'}{\lambda z}\right)\right|^2.
\end{equation}

In order to suppress grating lobes, we need to minimize the portion of $P_{det}$ due to grating lobes; for a uniform-reflectivity target, theoretically perfect suppression can be achieved for each grating lobe individually by aligning the peak of the RX grating lobe with a null of the corresponding TX grating lobe. See supplementary materials for further details. This peak-null alignment is controlled by the difference in grating pitches $\Lambda_{TX/RX}$, where for identical pitches (non-Verniered TX/RX pair) the peaks are always aligned and there is \emph{no} grating lobe suppression. This identical TX/RX case provides a reference value with which we can compare the Vernier design to determine the extra suppression provided by the Vernier transceiver.

The sidelobe nulls of a single beam are spaced at intervals of $1/W_{TX/RX}$ in the angular domain. To align the peak of every $m$\textsuperscript{th} RX grating lobe with the $n$\textsuperscript{th} null of each corresponding ($m$\textsuperscript{th}) TX grating lobe then requires

\begin{equation}
    m\left(\frac{1}{\Lambda_{RX}}-\frac{1}{\Lambda_{TX}}\right) = \frac{i}{W_{TX}},\quad m,n\in \mathbb{Z}
\end{equation}

\noindent which is automatically satisfied when $W_{TX} = W_{RX}$. Considering as an example the case of $N_{TX}=N_{RX}+k$, such that $k$ is the number of rows different between the TX and RX arrays, it can be seen that the first ($m=1$) grating lobe pair has the RX peak aligned with the $n=k$ null, the second ($m=2$) RX grating lobe is aligned to the $n=2k$ null, and so on. Notably, while the detected power is identically null for each grating lobe pair in isolation regardless of the value of $k$, the grating lobe suppression will not be perfect in realistic situations with non-uniform reflectivity and finite targets. Larger values of $k$ increase grating lobe suppression in realistic scenarios by further separating the grating lobe peaks.

This choice of geometry, the corresponding radiation patterns, and detected power is shown in Fig.~\ref{fig:theory}. In Fig.~\ref{fig:theory}\textbf{a} we depict the restricted geometry, identical OPA widths but where the RX OPA has $k$ fewer gratings than the TX OPA, for an example 6-grating TX OPA and $k=1$. The theoretical radiation patterns of the fabricated OPAs, discussed further in Sec.~\ref{sec:sopa}, are shown in Fig.~\ref{fig:theory}\textbf{b}. This design uses a 32-grating TX OPA with 16 {\textmu}m grating pitch and 6.5 {\textmu}m wide gratings, which results in grating lobes at $\pm5.5^\circ$. The TX OPA radiation pattern (and identical RX OPA pattern for the $k=0$, standard transceiver case) is shown in blue, while the RX OPA radiation pattern for a Vernier transceiver design with $k=1$, $k=2$ is shown in red and blue respectively. The Vernier transceiver suppresses returns from the grating lobes to avoid signal ambiguity, which is depicted in Fig.~\ref{fig:theory}\textbf{c-d} for broadside emission and plotted in terms of detected optical power. The point target case is shown in Fig.~\ref{fig:theory}\textbf{c}, where the $k=1$ and $k=2$ Vernier transceivers have peak grating lobe returns 8.6 dB and 17.7 dB lower than the standard transceiver ($k=0$). A uniform target which is exactly one spot width (Fig.~\ref{fig:theory}\textbf{d}) will have grating lobe returns 17 dB and 20.8 dB lower than the standard transceiver.

\begin{figure*}[t]
	\centering
	\includegraphics[width=\textwidth]{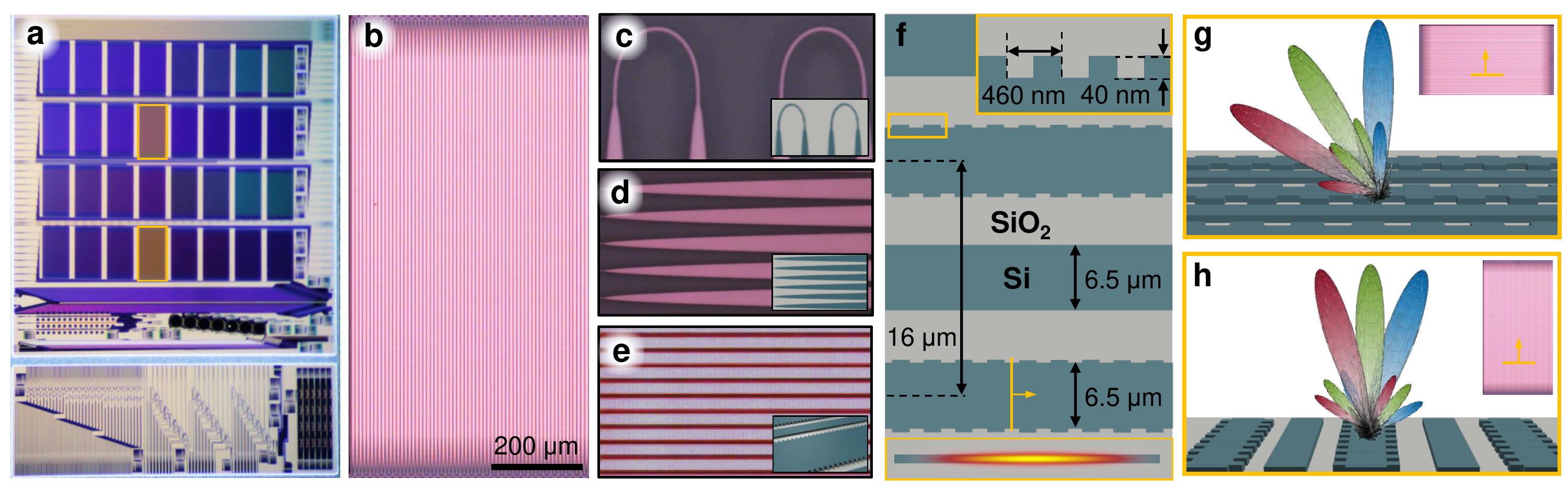}
	\caption{\textbf{Fabricated SOPAs used for the Vernier transceiver demonstration.}
	\textbf{a} Fabricated chip with the two SOPAs used for the Vernier transceiver ($k=1$) highlighted.
	\textbf{b} Optical micrograph of the TX SOPA.
	\textbf{c} Optical micrograph of the adiabatic bends; inset: render.
	\textbf{d} Optical micrograph of the tapers; inset: render.
	\textbf{e} Optical micrograph of the grating-waveguides and flyback waveguides; inset: schematic render.
	\textbf{f} SOPA layout with design dimensions (grating teeth not to-scale). Insets: (top right) grating teeth parameters; (bottom) waveguide cross-section with fundamental mode.
	\textbf{g} Wavelength-steering along a grating; inset: gold bar denotes cross-section location.
	\textbf{h} Wavelength-steering orthogonal to the gratings; inset: gold bar denotes cross-section location.
	}
	\label{fig:sopa}
\end{figure*}

\subsection{Beam Steering with Vernier: Tracking and Sliding Vernier Steering}
\label{subsec:steer_general}

In addition to choosing the geometry of the TX and RX OPAs, there is also the question of steering a Vernier transceiver's beam. The main lobes must remain aligned, corresponding to ideally misaligned grating lobes enforced by the geometry. Considering first a plane wave emitting from an aperture radiating to an angle $\theta_x$ relative to broadside, the phase-function in a the aperture plane is simply $\exp\left[j\beta_xx\right]$ where $\beta_x = (2\pi/\lambda)\sin(\theta_x)$. For an OPA, with discrete emitters, we steer the main lobe to the same angle by sampling this phase function at the emitter locations. For the $k$th emitter the phase is then

\begin{equation}
    \phi_k=\beta_xk\Lambda
\end{equation}

\noindent and the phase difference between adjacent emitters is $\Delta\phi = \beta_x\Lambda$. Notably, because $\Lambda_{TX}\neq\Lambda_{RX}$ for our grating lobe suppression geometry, the phase difference between adjacent emitters is not identical between the TX and RX OPAs. Rather, the `phase rate' $\partial\phi/\partial x = \beta_x$ is preserved in order to keep the main lobes aligned so that the RX lobe `tracks' the corresponding TX lobe. As an alternative choice, one might maintain the same `phase step' (phase difference $\Delta \phi$) for both TX and RX OPAs. This choice forces the main lobes to `slide' into and out of alignment while the beam is steering such that two lobes are always aligned near broadside. We denote these two conditions as the tracking and sliding steering modes, both of which will be of interest in this paper for wavelength-steered OPAs. For an OPA with phase-shifter steering, only the tracking mode is desirable.

For the prescribed relation of grating pitches, we can write these two steering modes (tracking and sliding, respectively) as

\begin{equation}
\label{eq:scanning}
\begin{split}
    \Delta\phi_{RX}= &\frac{\Lambda_{RX}}{\Lambda_{TX}}\Delta \phi_{TX}\\
    \Delta\phi_{RX}= &\Delta \phi_{TX}.
\end{split}
\end{equation}

These two steering modes are depicted in Fig.~\ref{fig:theory}\textbf{e-f}, respectively, for the $k=1$ case. Notably, the tracking mode maintains lobe alignment at all angles, enabling the full FOV to be recovered as desired in addition to grating lobe suppression. The sliding mode does \emph{not} expand the FOV beyond the grating lobe spacing, but does suppress grating lobes. Arbitrary steering modes, and discussion of the advantages of different steering modes in wavelength-steered OPAs, can be found in the supplementary materials.

\subsection{Wavelength Beam Steering}
\label{subsec:steer_wl}

With the generic beam steering functions determined, we are now interested in determining the more restrictive conditions of Vernier steering in wavelength-steered OPAs. The analysis of wavelength steering is linear for optical frequency $\omega$ and therefore we use the frequency rather than the wavelength to analyze wavelength-steering.

The simplest wavelength-steered OPAs uses an individual delay line input to each grating \cite{van2011two}, with each successive delay line incrementally longer than the previous my some length $\Delta L$. The phase of each grating is therefore tied to the wavelength via the delay line, allowing wavelength to be used to control the beam emission angle. For simplicity we assume all delay lines have identical effective and group indices $n_\text{eff}$, $n_g$ and we analyze the phase profile relative to a central frequency $\omega_0$ at which a lobe is emitted at broadside. For a frequency shift $\Delta\omega = \omega-\omega_0$ and small group velocity dispersion the phase step is then

\begin{equation}
    \Delta\phi(\Delta\omega) = \frac{n_g(\omega_0)\Delta L\Delta\omega}{c}.
\end{equation}

Noting that the quantity $n_g\Delta L/c$ is the group delay $\tau$ accumulated in the incremental length, we can write the phase step as simply

\begin{equation}
    \Delta\phi(\Delta\omega) = \tau\Delta\omega
\end{equation}

\noindent the product of the extra group delay and the frequency shift.

Using \eqref{eq:scanning} we can find the relation of incremental lengths required in the wavelength steering case for tracking and sliding, respectively, as

\begin{equation}
\begin{split}
    \Delta L_{RX} = &\frac{\Lambda_{RX}}{\Lambda_{TX}}\Delta L_{TX}\\
    \Delta L_{RX} = &\Delta L_{TX}.
    \end{split}
\end{equation}

For both of these relations, we could replace the incremental length $\Delta L$ with the incremental delay $\tau$ and find equally valid relations. Hence, since slow light waveguides can also be used to manipulate the steering (via the group index). One could also use the group index (with cross-section design) rather than length to produce the Vernier steering modes.

\begin{figure*}[t]
	\centering
	\includegraphics[width=\textwidth]{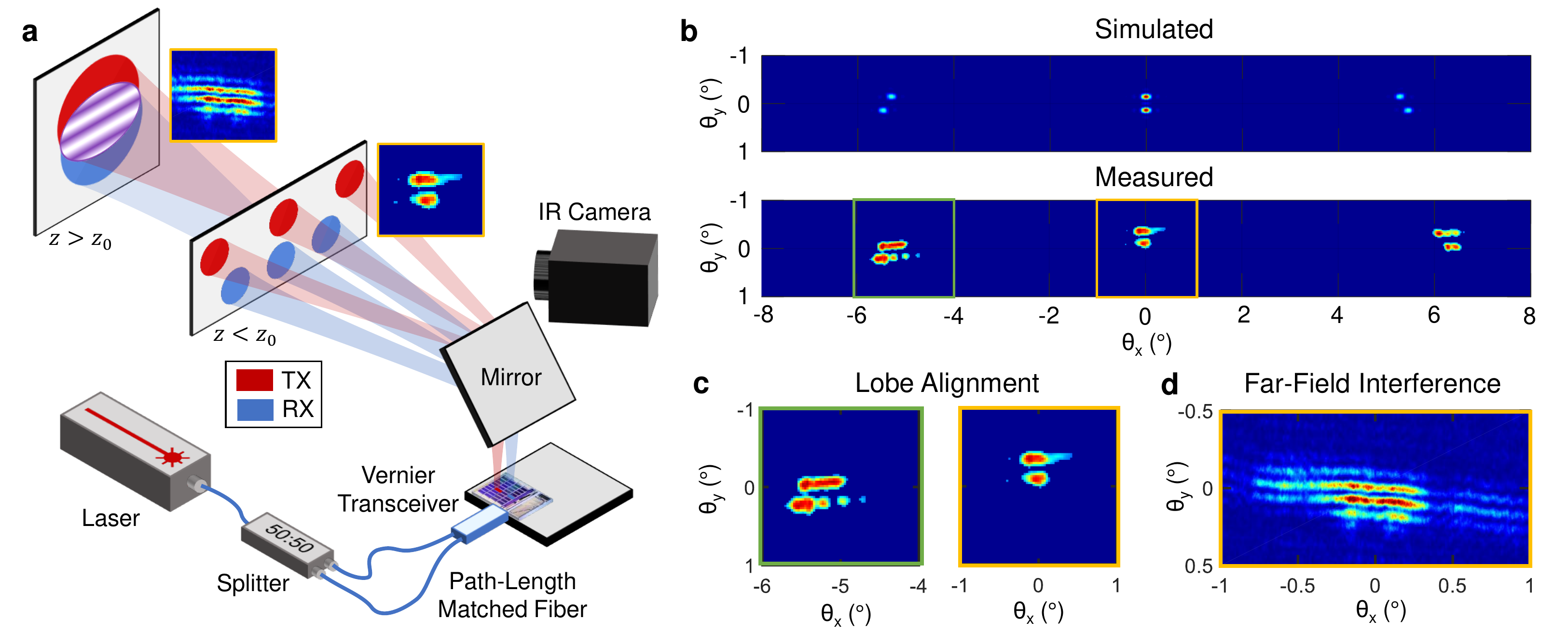}
	\caption{\textbf{Test setup for Vernier transceiver demonstration and measured radiation patterns.}
	\textbf{a} Test setup where a CW laser is split and passed to two SOPA simultaneously projecting two spot patterns to mid-field ($z< z_0$) and far-field ($z>z_0$) planes. The mid-field plane, placed at approximately the Rayleigh range $z_0$ of a single SOPA, is imaged using an IR camera and allows us to easily observe lobe alignment. The far-field plane is obtained using a Fourier lens and an IR detector placed in the Fourier plane, allowing accurate measurement of spot overlap. 
	\textbf{b} Simulated (top) and measured (bottom) radiation patterns of the two SOPAs in the mid-field plane. The lobe spacing is slightly uneven in the measured case due to a slight tilt of the measurement system.
	\textbf{c} Magnified view of both the misaligned grating lobes (green) and aligned main lobes (gold).
	\textbf{d} Measured far-field interference pattern of the main lobes demonstrating their overlap in the far-field.
	}
	\label{fig:setup}
\end{figure*}

\section{Serpentine Optical Phased Array}
\label{sec:sopa}

To demonstrate a Vernier transceiver, we use a 2D wavelength-steered OPA design demonstrated previously \cite{dostart2020serpentine} which we call a serpentine OPA (SOPA). We fabricated an array of SOPAs, shown in Fig.~\ref{fig:sopa}\textbf{a}, of which two (highlighted) are used to demonstrate a Vernier transceiver with approximately 2 mm center-to-center spacing. Optical micrographs of the TX SOPA can be found in Fig.~\ref{fig:sopa}\textbf{b-e}, and design dimensions are shown in Fig.~\ref{fig:sopa}\textbf{f}. The SOPA routes the output of a grating to the input of the adjacent grating, thereby using the entire aperture has a single delay line. This is results in 2D wavelength-steering shown schematically in Fig.~\ref{fig:sopa}\textbf{g-h} where coarse wavelength shifts steer the beam along the grating dimension and fine wavelength shifts steer the beam along the grating-orthogonal dimension.

For the demonstration in this paper we designed a pair of SOPAs, one TX SOPA and one RX SOPA, to create a Vernier transceiver. The TX SOPA uses 6.5 {\textmu}m wide waveguides for both the grating and flyback waveguides to achieve an ultra-low propagation loss (measured to be $<0.06$ dB/cm) \cite{dostart2020serpentine}. Our initial designs used a 16 {\textmu}m grating-to-grating pitch to ensure minimal cross-talk, resulting in a $5.5^\circ$ grating lobe spacing. A variety of grating variants were fabricated; for this demonstration we use a silicon-sidewall grating with 50\% duty cycle, 40 nm teeth, and 460 nm period (see Fig.~\ref{fig:sopa}\textbf{f}, inset). The TX SOPA has 32 gratings while the Rx SOPA has 31 gratings ($k=1$), both approximately 800 {\textmu}m long (sliding mode). A sliding mode was chosen to ensure a fully-populated FOV. See supplementary materials for additional discussion on implementation details for Vernier transceiver steering modes specific to SOPAs and design trade-offs.

\section{Results}
\label{sec:results}

To demonstrate the Vernier transceiver, with $k=1$ and a sliding mode, we \emph{transmit} from both TX and RX SOPAs simultaneously to measure their overlap patterns. The experimental setup is shown in Fig.~\ref{fig:setup}\textbf{a}, where the radiation pattern is measured either in a `mid-field' plane at approximately the Rayleigh range $z_0$ of a single aperture ($\sim1$ m) or in a far-field plane at a distance much greater than the Rayleigh range $z_0$ of the total, 3 mm long aperture. While the Rayleigh range of a single aperture can be easily accessed on a table top, the far-field plane of the composite aperture is approximately 20 m and we therefore use a Fourier lens to focused onto an IR detector to access the far-field plane.

The mid-field plane allows for easily distinguishable lobe alignment by simultaneously transmitting from both SOPAs, shown in Fig.~\ref{fig:setup}\textbf{b}. The grating lobe spacing, as predicted by theory, is approximately $5.5^\circ$. The lobes are slightly bleached by overexposure of the detector and we have applied a threshold to the image to suppress spurious noise. The increase in spot width over the simulated spot size is mainly a result of phase errors accumulated across the 6.4 cm path length of each SOPA. For additional details, see \cite{dostart2020serpentine}.

The mid-field plane allows one to see clearly see the alignment of the main lobes and misalignment of the grating lobes (Fig.~\ref{fig:setup}\textbf{c}). However, the far-field plane is needed in order to accurately measure the overlap of the two lobes, which requires transmission from only one SOPA at a time. The overlap of the main lobes transmitted from both SOPAs in the far-field is shown in Fig.~\ref{fig:setup}\textbf{d}, where the the high visibility and uniformity of the fringes demonstrates the high phase-coherence between the two SOPAs.

In this initial demonstration we did not measure LIDAR returns, so we calculate grating lobe suppression with a proxy metric, the overlap of the grating lobes. By transmitting from only the TX SOPA and measuring the far-field radiation pattern at each wavelength, and repeating this procedure for the RX SOPA, we measure the intensity distribution of each SOPA and calculate an overlap. Without direct access to the radiated \emph{field}, only the intensity, we therefore calculate the intensity overlap of the radiated lobes and compare to the theoretical case. This intensity overlap is simply $\int(I_{TX}I_{RX})^2dx/(\int I^2_{TX}dx\int I^2_{RX}dx)$. This allows measurement of the grating lobe suppression (by proxy) and demonstration of the sliding mode implemented here.

\begin{figure*}[t]
	\centering
	\includegraphics[width=1\textwidth]{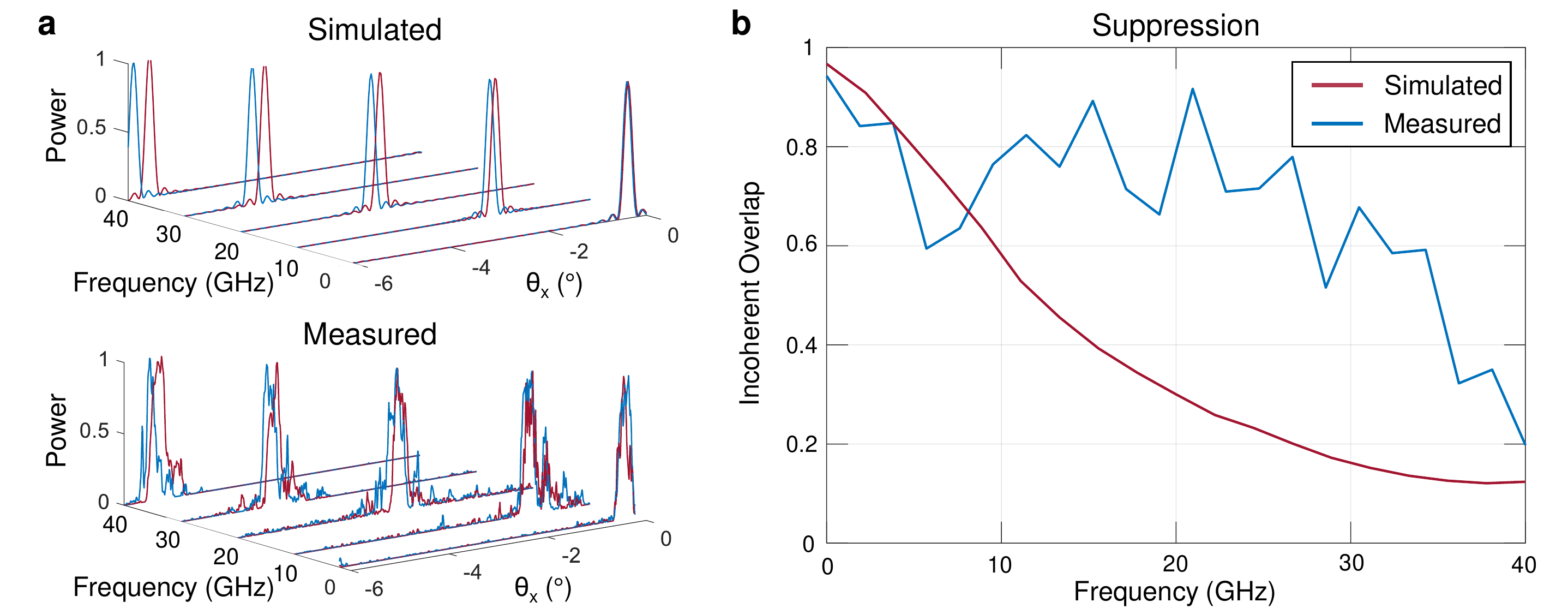}
	\caption{\textbf{Demonstration of sliding steering mode and grating lobe suppression with a Vernier transceiver.}
		\textbf{a} Simulated and measured radiation modes during a single scan along the grating-orthogonal ($x$) dimension. The 1D radiation modes are obtained by integration along the grating-dimension, and only 5 out of 24 measured cross-sections are plotted for clarity. Over the 41.3 GHz frequency range required to scan the $x$-dimension \cite{dostart2020serpentine}, the initially overlapping spots slide out of alignment resulting in lobe suppression.
		\textbf{b} Simulated (red) and measured (blue) lobe overlap, a proxy measure for lobe suppression. The main lobes, initially aligned for broadside emission/reception, are increasingly suppressed as the scan past the edge of the FOV (20.5 GHz) and maximally suppressed outside this central FOV where they become grating lobes (up to 80\%, theoretically 88\%).
	}
	\label{fig:results}
\end{figure*}

In Fig.~\ref{fig:results} we demonstrate the sliding mode and grating lobe suppression. The sliding steering mode, discussed further in Sec.~\ref{sec:theory}, ensures the main lobes are aligned at broadside but slide apart as they are scanned across the FOV. At the edge of the FOV, these two lobes have slide partially out of alignment; at the other edge of the FOV, two grating lobes are equally misaligned and will enter the FOV as the scan continues, becoming effectively the main lobes. As can be seen in Fig.~\ref{fig:results}\textbf{a}, the measured scanning pattern agrees well with predictions. While the imperfect radiation patterns partially obscure this sliding, the centroids of the spots are misaligned for the grating lobes (40 GHz). The desired suppression can be more easily seen by computing the intensity overlap for various frequency shifts, which agrees well with predictions for the main and grating lobes for the broadside emission case (main lobe at 0 GHz, grating lobe at 40 GHz). The increasing suppression with detuning demonstrates that the lobes slide out of alignment as they are scanned across the FOV, as desired. Notably, we measured an incoherent overlap as low as 20\%, close to the theoretical value of 12\%. This predicts a 6.4 dB grating lobe suppression (compared to 8.6 dB ideal suppression), to be verified in future work.

\section{Discussion}
\label{sec:discuss}

These initial results indicate a Vernier transceiver is a promising approach to suppressing the ambiguous signals associated with grating lobes, and with improved designs can increase the FOV beyond the grating lobe spacing. The measured lobe suppression indicates the current demonstration is already useful for improving signal fidelity with regards to erroneous contributions from grating lobes. Additionally, this ambiguity suppression does not incur any additional loss for the return signal. By avoiding radiating out the power to all angles as with aperiodic OPA approaches \cite{hutchison2016high}, the Vernier transceiver can fully recover the signal while also avoiding the white noise background which inevitably results from aperiodic approaches.

The presented results are, however, very preliminary. Further work is needed to fully demonstrate grating lobe suppression by either directly measuring the radiated field or detecting a back-scattered signal for an imaging or LIDAR measurement. Such a setup would directly measure the suppression of grating lobes due to the Vernier transceiver as compared to a standard transceiver, whereas here we measure a proxy value -- the incoherent overlap. Furthermore, higher grating lobe suppression can be achieved using a $k>1$ design, which will be the subject of later experiments. Notably, up to 18 dB of grating lobe suppression (21 dB for a large target) can be achieved with a $k=2$ design, increasing as $k$ increases. This is an order of magnitude higher than the theoretical suppression of 8.6 dB, and with the presented data indicating 6.4 dB of suppression it is reasonable to expect a $k=2$ design to achieve upwards of 15 dB suppression. The most important aspect of increasing grating lobe suppression, however, is improving spot quality so that lobe alignment/misalignment can better suppress erroneous returns. Future iterations of the SOPA will either include phase-shifters to correct for the fabrication variations which degrade the spot quality, or be smaller in size so as to be unaffected by these variations. Additionally, the low risk sliding design we demonstrate in this paper does not increase the FOV beyond the grating lobe spacing; future iterations will use Vernier transceivers with a tracking design so as to increase the FOV as discussed in Sec.~\ref{sec:theory}. A tracking mode is particularly easy to implement in a phase-shifter steered OPA, and we expect that future demonstrations of Vernier transceivers in that subset of OPAs will use the tracking mode directly.

This paper presents a theoretical evaluation and experimental demonstration of a Vernier transceiver for reducing signal ambiguity and increasing FOV. We detail the optimal geometry for maximum grating lobe suppression, and the relative phases of emitters required to achieve different steering patterns. In particular, we discuss the non-trivial question of implementing a Vernier transceiver in 2D wavelength-steered OPAs and fabricate an example Vernier transceiver using our SOPA design. By simultaneously or alternately transmitting from a pair of SOPAs, we demonstrate alignment of main lobes at broadside, ensuring high signal detection, and misalignment of the grating lobes, ensuring rejection of erroneous signals. The measured spot patterns indicate up to 6.4 dB of grating lobe suppression, close to the theoretical value of 8.6 dB. Further improvements to the demonstrated design could suppress grating lobes by nearly 20 dB, and can widen the FOV up to the theoretical limit (the radiation pattern of a single emitter). We expect that Vernier transceivers will be a useful system-level design enabling improved OPA-based LIDARs and imagers, allowing these integrated photonic technologies to compete with traditional system designs.

\vspace{8pt}

\noindent \textbf{\large Funding.}
U.S. Government; National Defense Science and Engineering Graduate Fellowship Program (NDSEG) (GS00Q14OADS139); National Science Foundation (NSF) (1144083); Packard Fellowship for Science \& Engineering (2012-38222).

\vspace{8pt}

\noindent \textbf{\large Acknowledgments.}
Portions of this work were presented at CLEO 2019 in the paper \textit{Vernier Si-Photonic Phased Array Transceiver for Grating Lobe Suppression and Extended Field-of-View}, AW3K.2 \cite{dostart2019vernier}.  Chip layout was carried out using an academic license of Luceda Photonics IPKISS.

\vspace{8pt}

\noindent See supplementary materials for supporting content.

\bibliographystyle{ieeetr}
\bibliography{biblio}

\end{document}


\maketitle

\begin{abstract}
This document provides supplementary information to “Verniered Optical Phased Arrays for Grating Lobe
Suppression and Extended FOV.”
The supplementary material includes additional theoretical assessment of the Vernier transceiver concept. In particular, optimal lobe suppression design for arbitrary apodizations and additional considerations for wavelength-steered phased arrays.
\end{abstract}

\section{Choice of Optimal RX Array}

While in the body of the paper we restricted the RX array to have few gratings and larger pitch than the TX array, and heuristically arrived at the condition to align a TX peak with an RX null, we show here that these conditions can be arrived at by considering the maximum grating lobe suppression. For a set of beams emitted by the TX array, and an equivalent set of beams comprising the detection pattern of the RX array, we can evaluate the overlap integral equivalent to the (power) detection efficiency in angular space as

\begin{equation}
    \eta = \left|\int_{-\infty}^{\infty} df_x F_{RX}(f_x)F_{TX}(f_x)\right|^2.
\end{equation}

Choosing to look pairwise at each TX/RX beam pair, from a rectangular aperture, we require the overlap integral to evaluate to null. The offset $\Delta f_x(k)$ between the $k$\textsuperscript{th} beam pair centers is determined by their respective grating pitches for broadside emission as

\begin{equation}
    \Delta f_x(k) = k\left(\frac{1}{\Lambda_{TX}}-\frac{1}{\Lambda_{RX}}\right)
\end{equation}

\noindent and the overlap integral for the $k$\textsuperscript{th} beam pair can therefore be written in angular space as

\begin{equation}
    \int_{-\infty}^{\infty} df_x \text{sinc}\left(W_{TX}(f_x-\Delta f_x(k))\right)\text{sinc}\left(W_{RX}f_x\right)=0
\end{equation}

\noindent where we have neglected scaling factors, set the RX beam to be centered at $f_x=0$, and identified the aperture widths as $W_i=N_i\Lambda_i$.

By choosing to set this integral to 0, rather than minimize the overlap function $F_{RX}F_{TX}$, we are implicitly assuming our desired target is resolved by the far-field spot. In this case all the transmitted power intercepts the target and is reflected back to create the signal, so we desire to maximize the power received from the main lobe and minimize the power received from other lobes (minimize the overlap integral). For unresolved targets, the signal is proportional to beam intensity rather than power and in that case we would minimize the received intensity from other lobes (minimize the overlap function).

Noting that this integral can be evaluated as a convolution of the form

\begin{equation}
    \left[ \text{sinc}\left(W_{TX}f_x\right)\ast\text{sinc}\left(W_{RX}f_x\right)\right]_{\Delta f_x(k)} = 0
\end{equation}

\noindent we Fourier transform the convolution to get a multiplication of rects, the two aperture functions. We therefore have

\begin{equation}
    \left[ \mathcal{F}^{-1}\left\{\text{rect}\left(\frac{x}{W_{TX}}\right)\text{rect}\left(\frac{x}{W_{RX}}\right)\right\}\right]_{\Delta f_x(k)} = 0.
\end{equation}

\noindent However, this can be easily simplified to a single rect with the width of the \emph{smaller} aperture. Therefore, for a given TX array width, there is no effect on the overlap (grating lobe suppression) by making the RX array \emph{larger} than the TX array. However, if the RX array is smaller than the TX array the overlap will be affected (negatively, as we will show). This will motivate choosing an RX array the same size as the TX array.

Denoting the effective aperture width as $W=\text{min}[W_{TX},\,W_{RX}]$, the null overlap condition is

\begin{equation}
    \text{sinc}\left(W\Delta f_x(k)\right) = 0.
\end{equation}

\noindent This condition sets a restriction on the beam center offset $\Delta f_x$ which can be understood more intuitively as choosing to align the peak of one radiation pattern with the null of the other. Specifically, we require for $k\neq0$ (i.e. all grating lobes)

\begin{equation}
    W\Delta f_x(k) = m,\quad m \in \mathbb{Z} : m \neq 0
\end{equation}

\noindent or equivalently

\begin{equation}
    W\left(\frac{1}{\Lambda_{TX}}-\frac{1}{\Lambda_{RX}}\right) = l,\quad l\in\mathbb{Z}:l\neq0.
\end{equation}

Notably, for $N_{TX}-N_{RX}=\Delta N$, and choosing identically sized TX and RX arrays $W=W_{TX}=W_{RX}$ (thereby setting $\Lambda_{RX} = \Lambda_{TX}N_{TX}/N_{RX}$), this null overlap is guaranteed. This can be seen by evaluating the null overlap condition which is

\begin{equation}
\label{eq:null_overlap_condition}
    W\left(\frac{1}{\Lambda_{TX}}-\frac{1}{\Lambda_{RX}}\right) = \Delta N
\end{equation}

\noindent recovering the solution we arrived at in the body of the paper heuristically.

We note here that smaller aperture widths, which determine the effective overlap width $W$, increase the minimum pitch difference between the TX and RX arrays required to reach a given null of the sinc. Other considerations, in particular power loss to unused grating lobes, motivate small pitches for \emph{both} the TX and RX arrays. We therefore desire to use the smallest pitch possible for both arrays, and for one array we sacrifice the pitch in order to achieve this peak null alignment for grating lobe suppression. We should then avoid worsening this sacrifice by decreasing the RX array width below the TX array width, and there is no benefit gained with regard to grating lobes suppression by making the RX array larger than the TX array. This consideration therefore motivates making both arrays the same size, again arriving at a conclusion which was intuitively arrived at the in the main body of the paper.

\section{Grating Lobe Nulling with Arbitrary Apodization}

Regarding other apodization patterns, in general this method of "perfect" grating lobe suppression using the Vernier method can be applied to any apodization function with few restrictions. For the purposes of the derivation we will restrict both TX and RX arrays to have the same apodization, but the same method is applicable for any combination of apodizations so long as the correlation of their Fourier transforms has a zero-crossing.

For identical TX and RX array apodizations, to achieve ideal grating lobe suppression we require an apodization which, when squared, has a Fourier transform with at least one zero-crossing. More specifically, we require that the aperture amplitude (without restrictions on phase) be a function $f(x)$ such that $\mathcal{F}^{-1}\{f^2(x)\}=0$ for some spatial frequency $f_x=a$.  In order to suppress \emph{all} grating lobes simultaneously, the nulls of the Fourier transform should be periodic and aligned to $f_x=0$. Notably, the vast majority of windowing/apodization functions meet these criteria \cite{oppenheim1999discrete}, with a few exceptions such as the Hanning-Poisson window \cite{harris1978use}. Any window function which is not `smooth', or continuous at all derivatives throughout the window, will have sidelobes in the Fourier domain \cite{harris1978use}. The set of functions which are smooth and have compact support are classified by mathematicians as `mollifiers' \cite{friedrichs1944identity}. However, this smoothness is a necessary but not sufficient criterion to avoid sidelobes; some functions within this set will still have sidelobes, such as the `standard' mollifier \cite{evans2010partial}.

Taking the null overlap condition for a general far-field distribution $F(f_x)=\mathcal{F}\{f(x)\}$ we have

\begin{equation}
    \int_{-\infty}^{\infty} df_x F(f_x)F(f_x-\Delta f_x)=0.
\end{equation}

\noindent Noting again that this equation is identical to an auto-convolution, we can rewrite the equation as

\begin{equation}
    \left[F(f_x)\ast F(f_x)\right]_{\Delta f_x}=0
\end{equation}

\noindent or, in the Fourier domain,

\begin{equation}
    \left[\mathcal{F}^{-1}\left\{f^2(x)\right\}\right]_{\Delta f_x}=0.
\end{equation}

The suppression condition is easily obtained by choosing $\Delta f_x$ as the location of a null of the auto-correlation of $F(f_x)$. For different apodizations, we equivalently require 

\begin{equation}
    \left[F(f_x)\star G(f_x)\right]_{\Delta f_x}=0.
\end{equation}

If this condition is enforced for the first grating lobe pair, the null periodicity about the symmetry plane ensures that all higher order grating lobes are also suppressed.

\section{Sliding Steering Mode}

To see the broadside alignment property of the sliding steering mode, consider a Vernier transceiver with the main lobes perfectly aligned at broadside $\phi_k = 0$ with the sliding mode applied $\Delta \phi_{TX}=\Delta\phi_{RX}$. If we were to steer the the main lobe of the transmit array to the location of its first ($k=1$) grating lobe, corresponding to a spatial frequency $\beta_x = 2\pi/\Lambda_{TX}$, the receive array main lobe will be pointing to its first grating lobe at $2\pi/\Lambda_{RX}$ because $\beta_x\Lambda_{RX}=\beta_x\Lambda_{TX}$. We can then see that, during this steering process, the two main lobes become misaligned. We can also see that the other first order grating lobes at $k=-1$ are now aligned and pointing at broadside. Because the beam steering is linear in terms of spatial frequency, it can be found that at halfway through this steering the main lobes (halfway to the $k=1$ grating lobe location) are partially misaligned and indistinguishable from the $k=-1$ grating lobes (halfway to broadside). It is therefore clear that the sliding mode always aligns two lobes (which are the effective main lobes) near broadside where the alignment region is simply the grating lobe spacing.

\section{Wavelength Beam Steering with Vernier}
\label{subsec:steer_wl_vernier}

We can now analyze both steering modes in the context of wavelength steering. Notably the incremental delay length can be different for transmit and receive arrays, and as both arrays need to operate at the same wavelength the steering mode restricts the incremental delay lengths.

For the tracking mode, we require $\Delta\phi_{TX}/\Lambda_{TX}=\Delta\phi_{RX}/\Lambda_{RX}$ and the incremental delay lengths are related as

\begin{equation}
\label{eq:tracking}
    \Delta L_{RX} = \frac{\Lambda_{RX}}{\Lambda_{TX}}\Delta L_{TX}.
\end{equation}

For the sliding mode we require $\Delta\phi_{TX}=\Delta\phi_{RX}$, corresponding to identical incremental delay lengths

\begin{equation}
\label{eq:sliding}
    \Delta L_{RX} = \Delta L_{TX}.
\end{equation}

For both of these relations, we could replace the incremental delay $\Delta L$ with the incremental delay $\tau$ and find equally valid relations. This identify will prove useful for the SOPA case examined in the following section.

\subsection{Design Considerations for Steering Modes in SOPAs}

The SOPA design presents a unique consideration for Vernier beam steering, as the emitters no longer have fully independent delay lines -- the emitter length factors into the incremental delay. Using the temporal delay interpretation of steering, we can write the phase step for the SOPA design as

\begin{equation}
\begin{split}
    \Delta\phi(\Delta\omega) & = \tau\Delta\omega\\
    & = \Delta\omega\left(\tau_g+\tau_f+2\tau_\text{bend}\right)
\end{split}
\end{equation}

\noindent where $\tau_g$ is the delay in the grating (emitter), $\tau_f$ is the delay in the flyback waveguide, and $\tau_\text{bend}$ is the delay in one taper-bend-taper structure, so that $\tau_g+\tau_f+2\tau_\text{bend}$ is one full row-to-row delay.

For simplicity we consider only changing $\tau_f$ and $\tau_g$ as the bend and taper designs are significantly constrained. Desiring to make a design robust to fabrication variations, one could aim to design the steering mode to be independent of the width and thickness of the waveguide. These parameters will affect the group index of a given fabrication run but unlikely to vary significantly between two adjacent SOPAs. One therefore expects that for a given grating cross-section and a (different) flyback cross-section there will be two unknown group indices $n_{g,g}$ (grating), $n_{g,f}$ (flyback) which are identical for TX and RX SOPAs. This leaves only the grating length $L_g$ as an unconstrained variable, as the connected nature of the geometry forces the flyback length to be set relative to the grating length for the desired straight waveguides. Changing the grating length does not change the scanning rate along the grating length, but does change the size of the spot in the far field along this dimension. For a set grating length of the TX SOPA, we therefore desire to minimally change the grating length of the RX SOPA for optimal spot overlap.

Under the restriction of equal flyback and grating lengths, the total delay is

\begin{equation}
    \tau = 2\tau_\text{bend}+\frac{L_g}{c}\left(n_{g,g} + n_{g,f}\right)
\end{equation}

\noindent allowing us to set the grating length according to the steering mode. As stated previously in Eqs.~\eqref{eq:tracking}-\eqref{eq:sliding}, the tracking and sliding modes respectively correspond to

\begin{equation}
    \tau_{RX} = \frac{\Lambda_{RX}}{\Lambda_{TX}}\tau_{TX}
\end{equation}
\begin{equation}
    \tau_{RX} = \tau_{TX}.
\end{equation}

While the sliding mode implies that both TX and RX SOPAs should have identical grating lengths, the tracking mode requires $\tau_{RX}>\tau_{TX}$ for $\Lambda_{RX}>\Lambda_{TX}$. In this case we need a longer grating to achieve the increased delay. An intuitive solution is found in the case where the delay in the bends can be neglected, in which case $L_{g,RX} = \Lambda_{RX}/\Lambda_{TX}L_{g,TX}$. Recalling our earlier suggested configuration of $N_{TX}\Lambda_{TX} = N_{RX}\Lambda_{RX}$ for Vernier nulling of adjacent sidelobes, we see that the same phase delay condition enforces that the total waveguide length of the transmit and receive arrays be identical $N_{TX}L_{g,TX}=N_{RX}L_{g,RX}$ when there is no delay in the bends.

\subsection{Comparison of Steering Modes}

It is worth making a few comments at this point on both Vernier steering modes which, upon further examination, will point to an optimized steering mode which outperforms either of these basic approaches.

Sliding mode has the property of guaranteeing that two lobes will be aligned at some angle within the hemisphere due to the quality of the two lobes "sliding" across each other within the FOV during a wavelength scan. However, for misaligned lobes, this FOV will not be centered on broadside. The two most-aligned lobes are within this off-broadside FOV and, within a single fast scan, will become perfectly aligned at some angle due to the sliding property. This perfect alignment angle is the center angle of the FOV. For a $k=1$ Vernier transceiver (see Eq.~\eqref{eq:null_overlap_condition}) an FOV scan will lead to a sliding of the transmitter spot fully over the receiver spot.

This contrasts with the tracking mode, where the relative alignment of every pair of lobes is locked in place throughout the wavelength scan. Notably the Vernier arrangement guarantees one pair of lobes to be nearly aligned, and the FOV is unrestricted by any alignment condition (unlike sliding mode). Additionally, because the alignment is locked in throughout the scan range, the effective FOV is simply the radiation pattern of a single grating which is centered on broadside.

While tracking mode is clearly superior in this regard, there is a price paid for the increased FOV and guaranteed alignment in 2D wavelength-steered arrays. A significant portion of the full hemispherical lobe alignment of tracking mode is spent outside the effective FOV. This fact highlights that we lose a portion of the wavelength scan by scanning over angles where we can neither transmit nor receive signal as the aligned lobes are directed outside the radiation pattern of a single grating. It is convenient at this point to consider that, for a given design with particular row-to-row delay, there is a set frequency shift required to move to the next spot along the fast axis. If the wavelength scan range is also set, then there is constant number of spots we can steer to. We can see then that the price of the guaranteed alignment and extended FOV of tracking mode is a reduction in the total number of usable, addressable spots. For sliding mode, when the perfect alignment angle is within the radiation pattern of a single grating, we are able to access all potential spots. The highest efficiency over the scan results when the alignment angle is in the middle (at the peak) of the single grating radiation pattern (element pattern in RF terminology), with efficiency decreasing towards the edges of the scan.

One additional difference between the two approaches should be noted, which is the set of spots that can be addressed. The use of 2D wavelength steering means that beam steering along both dimensions is not independent, any change in wavelength will steer the beam along both axes. The different scanning properties of tracking and sliding mode along the grating-orthogonal ($x$) direction then motivates design for different row-to-row phase accumulation rates. Ideally, the transceiver will steer to every spot within a 2D FOV without gaps. This requires that a fast scan across the effective FOV (limited by either the lobe sliding or transmit grating element radiation pattern) steers the spot by one spot width \emph{along} the grating dimension. Because the two steering modes have different effective FOVs, the different Vernier steering modes are optimized by different delay line lengths.

In order to benefit from the increased FOV of the tracking mode without `missing' spots along the slow axis, we need to increase the row-to-row delay relative to the sliding mode (decreasing the frequency shift needed to steer by one spot along the fast axis). This increases the number of spots within a full wavelength scan. In fact, for this optimal steering scenario, because the FOV is larger along the fast axis without changing the spot size, tracking mode can access more spots than sliding mode. The price is the smaller frequency bandwidth available for operations such as ranging \cite{dostart2020serpentine} and sub-spot imaging \cite{wagner2019super} which are bandwidth-limited to the fast scan frequency shift. For a design (row-to-row delay) optimized for tracking mode, the sliding mode implementation has an over-sampled FOV (overlapping spots). In the opposite situation (tracking mode using sliding design), there are gaps along the grating dimension between scans which cannot be accessed (too few spots).

\subsection{Optimal Phase Steering Modes for Controllable FOV Vernier}

Considering these two steering options, it can be seen that sliding mode suffers from limited FOV and potential misalignment whereas tracking mode throws away some scan wavelengths (in the case of the more efficient, but FOv-limiting wide grating element factor design) which are outside the FOV. This suggests that a hybrid steering mode which exactly scans over the maximum FOV without throwing away any scan wavelengths. Here, the maximum FOV achievable is that limited by by the radiation pattern of a single grating (the element pattern). The element pattern-limited FOV can be achieved by designing a Vernier which has lobes which slide apart by O(1 lobe width) over the desired FOV, rather than sliding apart over a lobe spacing as in sliding mode. Tracking mode can be seen as the case where the lobes do not slide apart, so we require a delay relation between the TX/RX OPAs which is between the sliding and tracking delay relations.

We denote the FOV for this controllable FOV Vernier in the same manner as the sliding mode: the angular range scanned over between perfect lobe alignment and the first peak-null lobe alignment. Considering the case of perfect lobe alignment at broadside, we need simply find the emitter phases which steers one lobe by an amount equal to the FOV and the other lobe to the corresponding null.

For this situation the TX SOPA main lobe is steered to the spatial frequency corresponding to the FOV width, and the RX SOPA main lobe is offset by half the lobe width. Denoting the angular FOV as $\theta_\text{FOV}$ the spatial frequency is $\beta_\text{FOV}=2\pi/\lambda\sin{\theta_\text{FOV}}$. The phase step between the emitters of the transmit and receive arrays are then

\begin{equation}
    \Delta\phi_{TX}=\beta_\text{FOV}\Lambda_{TX}
\end{equation}

\begin{equation}
    \Delta\phi_{RX}=\left(\beta_\text{FOV}-\frac{2\pi}{N_{RX}\Lambda_{RX}}\right)\Lambda_{RX}
\end{equation}

\bibliographystyle{ieeetr}
\bibliography{biblio}